\documentclass[twocolumn,aps,superscriptaddress,prl]{revtex4-1}
\usepackage{amssymb}
\usepackage{lipsum}
\usepackage[dvipdfmx]{graphicx}
\usepackage{color}
\usepackage{float}
\usepackage{braket}
\usepackage{amsmath}
\usepackage{mathrsfs}
\usepackage{ascmac}
\usepackage{appendix}
\usepackage{natbib}
\usepackage{here}
\usepackage{hyperref}
\newcommand{\ctext}[1]{\raise0.2ex\hbox{\textcircled{\scriptsize{#1}}}}
\usepackage{url}

\definecolor{red}{rgb}{1,0,0}
\usepackage{ulem}
\def\vector#1{\mbox{\boldmath $#1$}}
\begin{document}
\title{Fractal defect states 
in the Hofstadter butterfly}

\author{Yoshiyuki Matsuki}
\email{ymatsuki@het.phys.sci.osaka-u.ac.jp}
\author{Kazuki Ikeda}
\email{kikeda@het.phys.sci.osaka-u.ac.jp}
\author{Mikito Koshino}
\email{koshino@phys.sci.osaka-u.ac.jp}

\affiliation{Department of Physics, Osaka University, Toyonaka, Osaka 560-0043, Japan}

\begin{abstract}

We investigate the electronic properties in the Bloch electron on a square lattice with vacancies in the uniform magnetic field.
We show that a single vacancy site introduced to the system creates a defect energy level in every single innumerable fractal energy gap in the Hofstadter butterfly. The wavefunctions of different defect levels have all different localization lengths depending on their fractal generations, and they can be described by a single universal function after an appropriate fractal scaling. We also show that each defect state has its own characteristic orbital magnetic moment, which is exactly correlated to the gradient of the energy level in the Hofstadter diagram. Probing the spatial nature of the defect-localized states provides a powerful way
to elucidate the fractal nature of the Hofstadter butterfly.


\end{abstract}
\maketitle
\paragraph{Introduction}
%
The Hofstadter butterfly is the energy spectrum of Bloch electrons moving in a two-dimensional lattice under a uniform magnetic field, 
which is characterized by a nested fractal band structure \cite{Harper_1955,Harper_19551,PhysRev.134.A1602,Azbel,1976PhRvB..14.2239H,Wannier,WOR}.
It has been actively studied in condensed matter physics \cite{PhysRevB.19.6068,honeycomb,PhysRevB.46.12606,PhysRevB.52.14755,PhysRevLett.86.1062,PhysRevB.67.104505}, 
and also from a wide variety of perspectives including mathematics \cite{operator,dif,func,PhysRevB.91.245104,doi:10.1063/1.4998635,Ikeda:2017uce,Ikeda:2018tlz} and quantum geometry \cite{Hatsuda:2016mdw,2018arXiv180611092D}.
Experimentally, the evidence of the fractal nature of the Hofstadter spectrum was found in 
various systems, such as GaAs/AlGaAs heterostructures with superlattices \cite{Schl_sser_1996,PhysRevLett.86.147,geisler2004detection},
ultracold atoms in optical lattices \cite{Jaksch_2003,PhysRevLett.111.185301,PhysRevLett.111.185302}, 
graphene-based moir\'{e} superlattices \cite{moire,Hunt1427,Ponomarenko2013}, photons with the superconducting qubits \cite{Roushan1175} 
and one-dimensional acoustic array \cite{PhysRevLett.80.3232,Richoux_2002,ni}.

Currently, however, the experimental observation of the butterfly is mostly limited to the measurement of the spectral structure and 
the transport properties. Actually, richer fractal information is encoded in the wavefunctions of the Hofstadter butterfly, 
but it is generally considered to be difficult to access in experiments. 
The characteristic spatial property of each wavefunction is generally averaged out 
in the physical observables due to the summation over the Bloch momentum.

In this paper, we theoretically propose that the spatial structure in the Hofstadter system can be elucidated by introducing a point defect to the system. 
In an electron system under a magnetic field, generally, a point disorder potential gives rise to defect localized states in the energy gaps between Landau levels \cite{doi:10.1143/JPSJ.36.959,Ando1974,doi:10.1143/JPSJ.39.279,PhysRevB.23.4802,PhysRevB.29.3303,GREDESKUL1997223,PhysRevLett.97.236804}.
The effect of lattice defects on the Hofstadter spectrum was investigated in some past works \cite{PhysRevB.78.125402,slamo_lu_2012,PhysRevB.85.235414,PhysRevB.87.235404,PhysRevB.71.125310,honeycombdefect,Matsuki_2019,PhysRevB.101.245132}, 
and the in-gap defect levels were found at a certain magnetic flux \cite{PhysRevB.101.245132}.
However, it has not been clear how the self-similar nature is manifested in the defect localized states.


In this letter, we study the Hofstadter problem with vacancy defects in a square lattice to investigate the fractal properties of defect states. 
We show that a single vacancy site introduced to the system creates a defect energy level in every single innumerable fractal energy gap in the Hofstadter butterfly. We find that the wavefunctions of different defect levels have all different localization lengths depending on their fractal generations, and importantly, the localization length of any levels can be approximately described by a single universal curve after an appropriate fractal scaling.
We also find that the defect states are accompanied by an orbital magnetic moment due to rotating electric current, and its magnitude exactly coincides with the gradient of the energy gap in the Hofstadter diagram. 
These results give a new quantitative perspective on the spatial fractal nature of the Hofstadter butterfly,
and provide a powerful way to elucidate the fractal nature of the Hofstadter butterfly
by probing the defect states.


\begin{figure}[h]
  \centering
 \includegraphics[width=\linewidth]{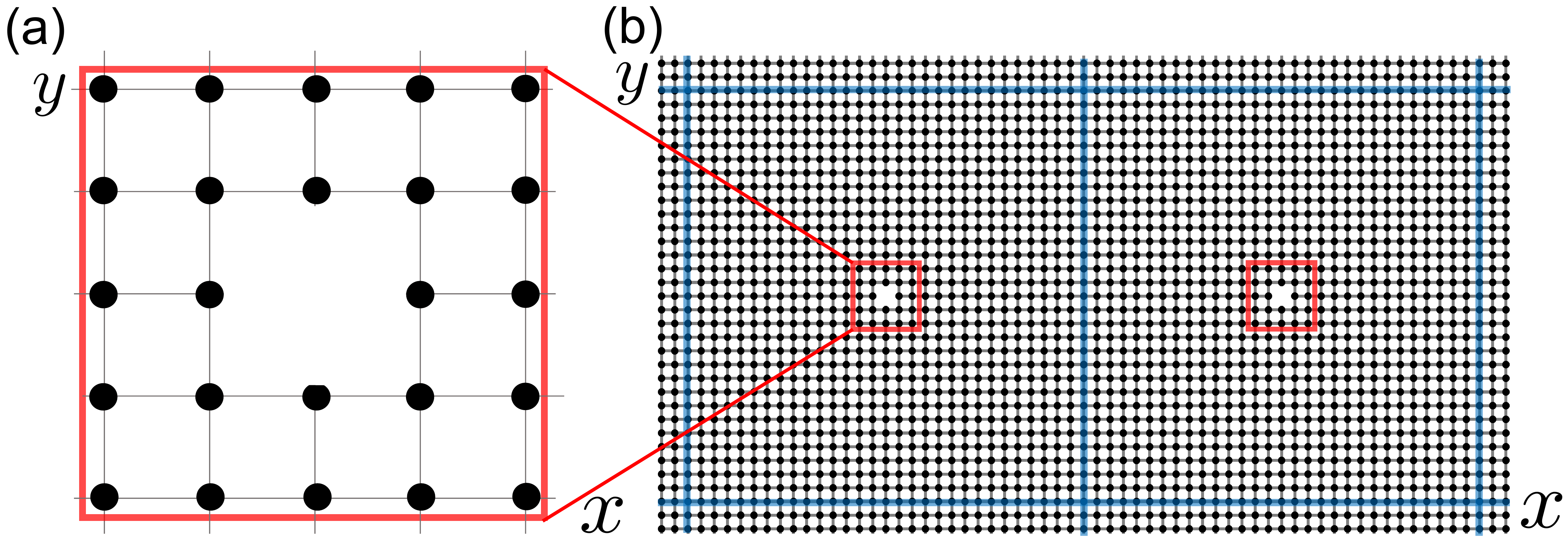}
  \caption{(a) Square lattice with a single point vacancy and (b) its periodic superlattice with $30\times30$ supercell}
 \label{fig:lat}
\end{figure}
\begin{figure*}[!ht]
\includegraphics[width=\linewidth]{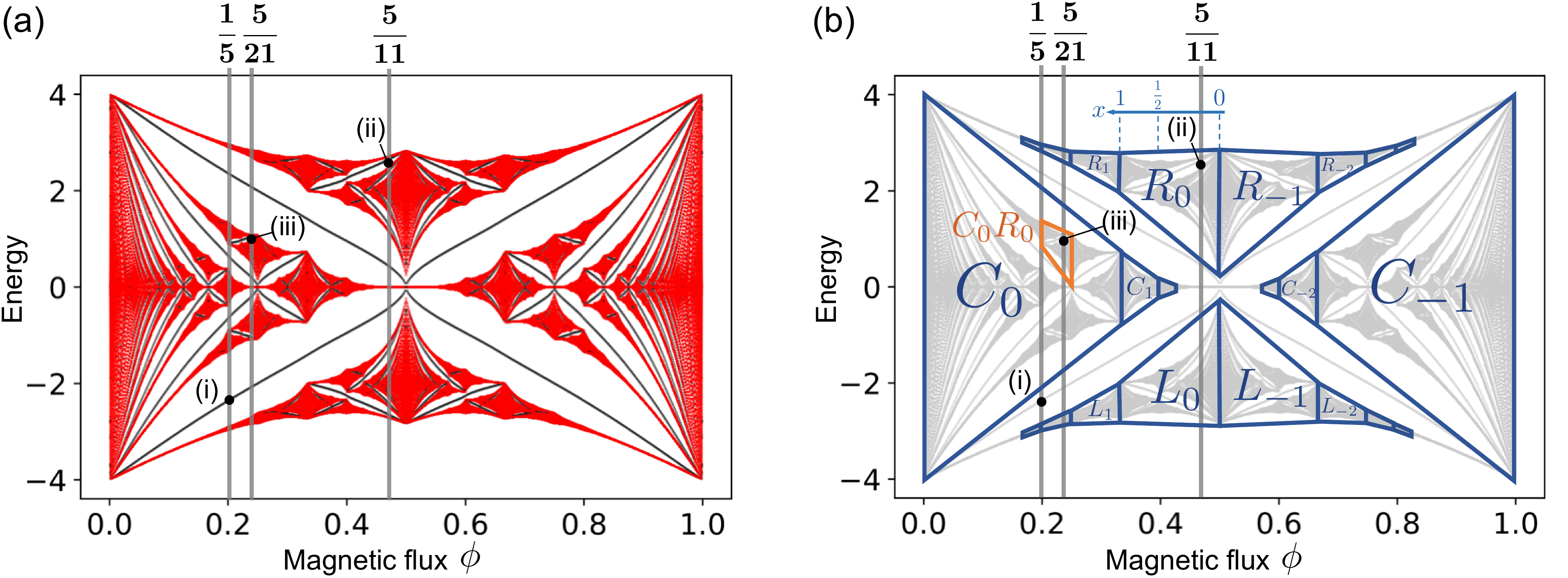}
\caption{(a) Energy spectrum of $30 \times 30$ superlattice with a single-site defect, which is plotted against the magnetic flux $\phi$.The red and black dots represent the bulks states and the defect-localized states, respectively. The labels (i), (ii) and (iii) correspond to the wavefunctions shown in Fig.\ \ref{fig:amplitude}. (b) Subcell decomposition of the Hofstadter butterfly (see the text). The states (i), (ii) and (iii) belong to the positive gradient principal gaps of the main spectrum, $R_0$ and $C_0 R_0$, respectively.} 
\label{fig:ingap}
\end{figure*}

\paragraph{Formulation}
We consider a square lattice with a single-site defect as illustrated in Fig.\ \ref{fig:lat}.
We assume that the system is periodic with $N\times N$ supercell and each supercell includes a single vacancy site.
The system is under a uniform magnetic field $B$ perpendicular to the system. Let $\phi=Ba^2/(h/e)$ be the number of magnetic flux quanta per a $1\times 1$ plaquette,
where $a=1$ is the spacing between the lattice points.
In what follows, we consider a single orbital tight-binding Hamiltonian,
\begin{equation}
    H=-t\sum_{\substack{\braket{m,n}\\ m,n \ne d}}\mathrm{e}^{i\theta_{mn}}c^{\dagger}_{m}c_{n},
\end{equation}
where $t \,(>0)$ is the hopping parameter, $\braket{m,n}$ is a pair of the nearest neighbor sites, $c^{\dagger}_{n}(c_{n})$ is the creation (annihilation) operator at site $n$, $d$ is the site of defects, $\theta_{mn}=-(e/\hbar) \int_{n}^{m} \vector{A}\cdot d\vector{\ell}$ is the Peierls phase \cite{Peierls1933},
and ${\vector{A}}=(0,Bx,0)$ is the vector potential.
We take $t=1$ throughout this letter. In a perfect lattice without a defect under a rational magnetic flux $\phi=p/q$ ($p, q$: coprime integers), the energy band splits into $q$ subbands \cite{1976PhRvB..14.2239H}.

The number of total magnetic fluxes penetrating an entire supercell is $\Phi = N^2 \phi$. For $\Phi=P/Q$ with co-prime integers $P$ and $Q$, the eigenstates of the system can be taken as magnetic Bloch states, which satisfy the following conditions \cite{PhysRev.134.A1602,RevModPhys.82.1959, PhysRevB.85.195458}:
\begin{eqnarray}
\psi(\vector{r}+\vector{L}_{1})&=&e^{ik_{x}L_{1}} e^{-i2\pi e BL_{1}y}\psi(\vector{r}),\\
\psi(\vector{r}+\vector{L}_{2})&=&e^{ik_{y}L_{2}}\psi(\vector{r}),
\end{eqnarray}
where $\vector{L}_{1} =(QNa,0)$ and $\vector{L}_{2}=(0,Na)$ are the primitive lattice vectors of the magnetic unit cell.
Then the eigen-energies and eigen-wavefunctions can be obtained by diagonalizing $QN^2\times QN^2$ Hamiltonian matrix.
We also perform similar calculation and analyses for a honeycomb lattice with periodic vacancies,
which is presented in Supplementary Information.

\paragraph{Fractal defect states}
Fig. \ref{fig:ingap}(a) shows the energy spectrum of $30 \times 30$ superlattice with a single-site defect, 
plotted against the magnetic flux $\phi$.
The red and black dots represent the bulk states and the defect-localized states, respectively.
Here the defect-localized states are identified by the condition that the wave amplitude within seven-site distance from the defect point
is more than $60\%$ of the total amplitude. 
We observe that a defect level exists in every single gap,
indicating that the spectrum of the defect states inherits the nested fractal structure of the Hofstadter butterfly.





\begin{figure}
    \centering
    \includegraphics[width=\linewidth]{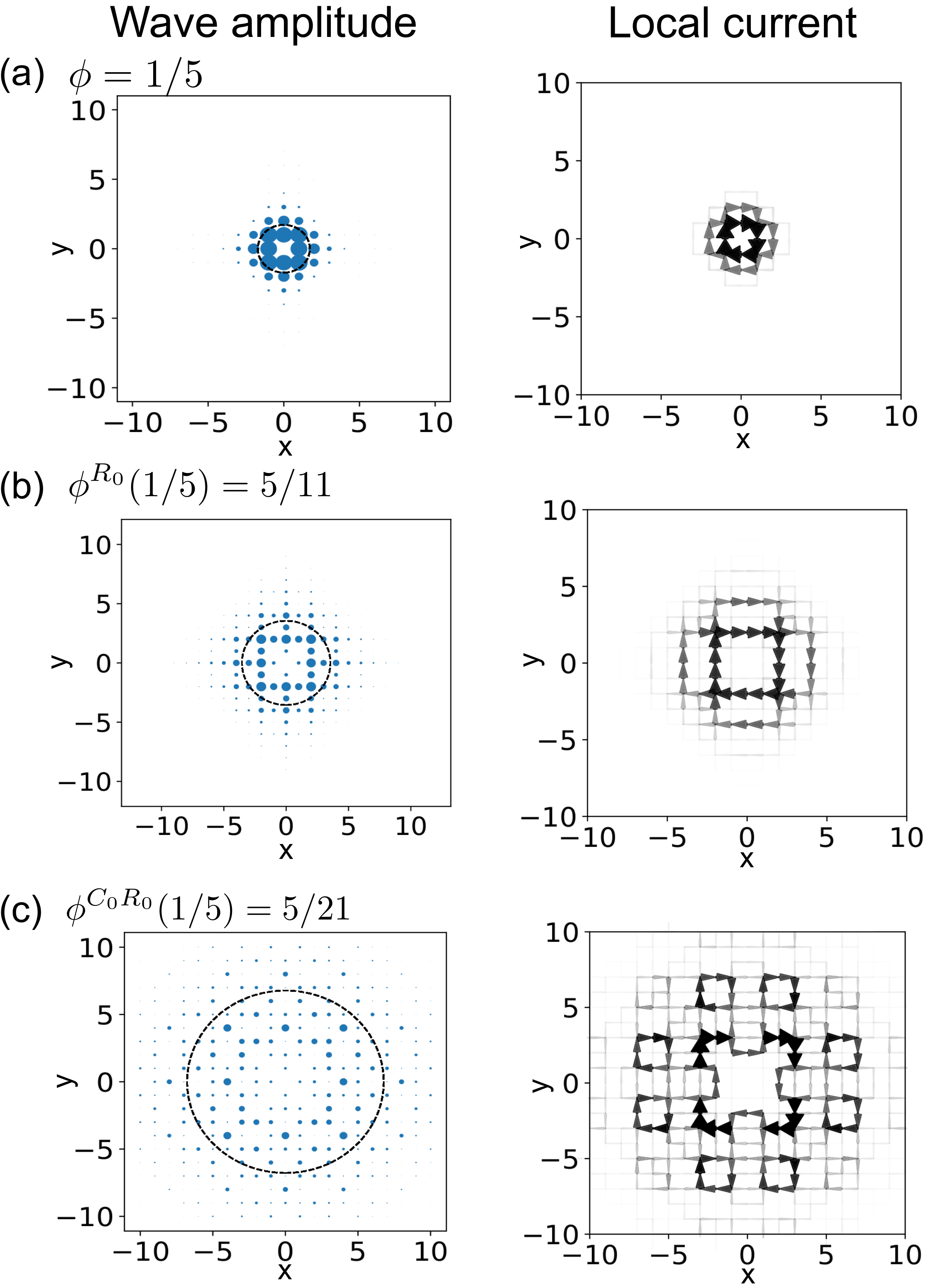}
    \caption{
Distribution of the wave amplitude (left) and the local electric current (right)
of the defect states (i), (ii) and (iii), which are indicated in Fig.\ \ref{fig:ingap}.
The dots size in the left figures represents the magnitude of wavefunction, and thickness and color depth of the arrows in the right figures represent the local current amplitude.}
    \label{fig:amplitude}
    \end{figure}

The left panels in Fig.\ \ref{fig:amplitude} represent the squared wavefunctions of defect levels (i), (ii) and (iii)  in Fig.\ \ref{fig:ingap}, which are taken from different minigaps of the Hofstadter butterfly.
Here the eigenstates are calculated for $40\times 40$ superlattice, where the overlap between defect states in neighboring unit cells is sufficiently small.
We clearly observe that the defect-state wavefunctions all localize around the vacancy, while their characteristic length scales are completely different.
Actually, as shown in the following, the localization length of the defect states is a good quantitative indicator of the fractal generation of the minigap.

To demonstrate this, we first introduce an addressing rule for the fractal structure \cite{1976PhRvB..14.2239H,PhysRevB.28.6713}. As shown in Fig.\ \ref{fig:ingap}(b), the whole energy spectrum in $0 \leq \phi \leq 1$ (referred to as the main spectrum in the following) can be divided into left, right, center subcells,
which are labelled by $L_n$, $R_n$ and $C_n$ $(n\in\mathbb{Z})$, respectively. The gap structure of each subcell plotted against the local variable $x\, (0 \leq x \leq 1)$ 
 is identical to that of the main spectrum plotted against the magnetic flux $\phi\, (0 \leq \phi \leq 1)$.
The global variable $\phi$ and the local variable $x$ for the subcell $X_n\,(X=L,R,C)$ 
are related by $\phi = \phi^{X_n}(x)$, where
\begin{align}
&\phi^{L_n}(x) = \phi^{R_n}(x) = (n+x+2)^{-1}, \label{eq_phi_fractal1} \\ 
&\phi^{C_n}(x)  =  [2+1/(n + x)]^{-1}. \label{eq_phi_fractal2}
\end{align}
In Fig.\ \ref{fig:ingap}(b), we present an axis of the local variable $x$ for $R_0$ subcell,
where $x=0,1/2,1$ correspond to $\phi=1/2,2/5,1/3$, respectively.
The main spectrum can also be regarded as a single subcell, where the local variable is the magnetic flux itself, i.e., $\phi = x$.
By repeating this addressing scheme, we can define subcells in higher generations.
For instance, $X_{m}Y_{n}$ refers to subcell $Y_{n}$ in subcell $X_{m}$ in the main spectrum.
The relation between the local variable $x$ for subcell $X_{m}Y_{n}$ and the global variable $\phi$
is given by $\phi= \phi^{X_m}(\phi^{Y_n}(x)) (\equiv \phi^{X_mY_n}(x))$.

For each subcell, we define the positive (negative) principal gap as the diagonal gap running from the lower (upper) left corner to the upper (lower) right corner of the subcell plotted against $\phi$. Any gap in the Hofstadter diagram can be uniquely identified as the positive or negative principal gap of a specific subcell or of the main spectrum. 


\begin{figure}[h]
\includegraphics[width=\linewidth]{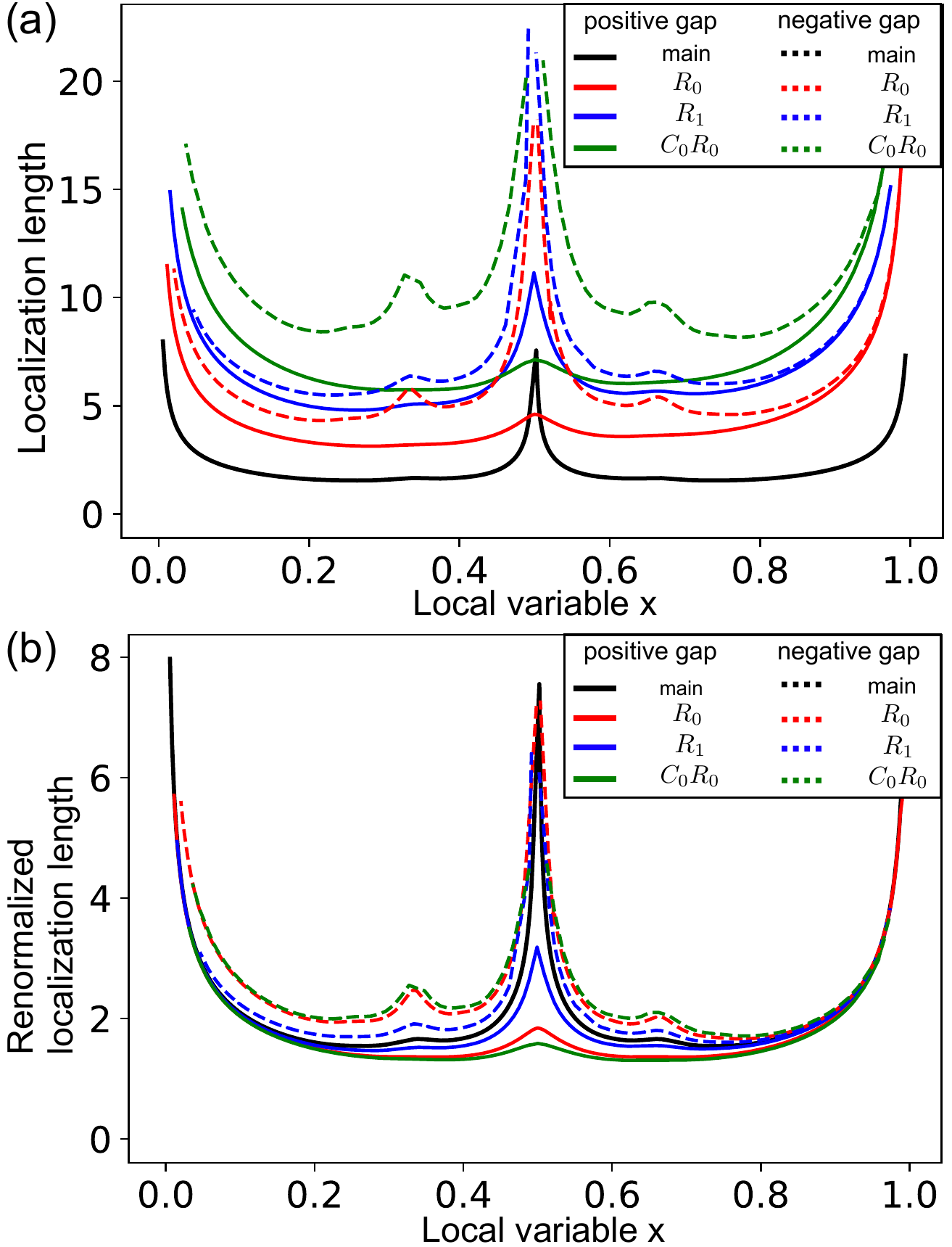}
\caption{(a) Localization lengths of different subcells, $\xi, \xi^{R_{0}}, \xi^{R_{1}}, \xi^{C_{0}R_{0}}$, 
and (b) their renormalized values $\xi, \phi^{R_{0}}\xi^{R_{0}}, \phi^{R_{1}}\xi^{R_{1}}, \phi^{C_{0}R_{0}}\xi^{C_{0}R_{0}}$ as functions of the local variable $x$.}
    \label{fig:fractal}
\end{figure}

In the following, we compare the localization lengths of the corresponding defect states in the principal gaps of different subcells, which share the same local variable $x$.
The defect levels (i), (ii) and (iii) in Figs.\ \ref{fig:ingap} and \ref{fig:amplitude}  are actually taken from the positive principal gaps of
the main spectrum, $R_0$ and $C_0 R_0$, respectively, with the same local variable $x=1/5$. 
The global variables $\phi$ for (i), (ii) and (iii)
are $1/5$, $\phi^{R_0}(1/5) = 5/11$ and $\phi^{C_0}(\phi^{R_0}(1/5))=5/21$, respectively.
For each state, we define the localization length $\xi$ by $\xi^2 = \sum_i |\mathbf{r}_i-\mathbf{r}_0|^2 |\psi(\mathbf{r}_i)|^2$,
where $\psi(\mathbf{r}_i)$ is the wave amplitude at site $\mathbf{r}_i$, and $\mathbf{r}_0$ is the vacancy position.
Here we show that the ratio of $\xi$'s of different subcells is approximately equal to the ratio of the denominators of $\phi$ of those states. For the states (i), (ii) and (iii) in Fig.\ \ref{fig:amplitude}, for instance, this claims that the ratio of the $\xi$'s of the three states is $5 : 11: 21$. Indeed, it approximates the ratio of the numerically calculated values $\xi=1.716, 3.547, 6.774$, respectively
(indicated by radii of circles in Fig.\
\ref{fig:amplitude}).
The reason for this scaling rule can be understood by considering an ideal system without defects.
Specifically, the Schr\"{o}dinger equation for the ideal square lattice with $\phi=p/q$ is reduced to a
one-dimensional Harper's equation with the spatial period of $q$ \cite{Harper_19551, 1976PhRvB..14.2239H}.
The period $q$ works as the reference length scale to compare the wavefunctions in different fractal levels;
for example, an eigenstate of Harper's equation at $\phi=1/5$ and the corresponding state of $R_0$ cell at $\phi = \phi^{R_0}(1/5) = 5/11$ have similar structures with length scales of $5 : 11$. Naturally, the defect states take over the same scaling feature.

Let $\xi(\phi)$ be the localization length of the defect state in the principal gap of the main spectrum at the flux $\phi$,
and $\xi^{X_n}(x)$ be that of subcell $X_n$ at local variable $x$.
According to the argument above, we have the relation $\xi(p/q)/q = \xi^{X_n}(p/q)/D[\phi^{X_n}(p/q)]$, where 
$D[\phi]$ is the denominator of $\phi$.
Using Eqs.\ (\ref{eq_phi_fractal1}) and (\ref{eq_phi_fractal2}), this immediately leads to the relation between $\xi$ and $\xi^{X_n}$,
\begin{eqnarray}
&\xi(x) =  \phi^{R_{n}}(x) \xi^{R_{n}}(x) =  \phi^{L_{n}}(x) \xi^{L_{n}}(x),\\
&\xi(x) =\displaystyle \frac{\phi^{C_{n}}(x)}{n+x}\xi^{C_{n}}(x),
\end{eqnarray}
which is a key finding of this work. Note that, although the denominator $D[x]$ is not a continuous function of $x$,
the scaling ratio $\xi^{X_{n}}(x) /\xi(x) $ is a continuous function of $x$.
Similarly, the relation for higher fractal generations can be obtained from 
 $\xi(p/q)/q = \xi^{X_mY_n\cdots}(p/q)/D[\phi^{X_mY_n\cdots}(p/q)]$.
For $C_0 R_0$ subcell, for instance, it gives $\xi(x) =  \phi^{C_0R_0}(x) \xi^{C_0R_{0}}(x)$.

In Fig.\ \ref{fig:fractal} , we plot (a) the localization lengths
$(\xi, \xi^{R_{0}}, \xi^{R_{1}}, \xi^{C_{0}R_{0}})$ 
and (b) the renormalized values $(\xi, \phi^{R_{0}}\xi^{R_{0}}, \phi^{R_{1}}\xi^{R_{1}}, \phi^{C_{0}R_{0}}\xi^{C_{0}R_{0}})$ 
as functions of the local variable $x$. Here the solid and dashed curves represent the positive and negative principal gaps, respectively.
The two curves are identical for the main spectrum due to the electron-hole symmetry.
We see that the renormalized localization lengths [Fig.\ \ref{fig:fractal}(b)] quantitatively match in a wide range of $x$. 
 The $\xi(x)$ diverges at $x=0$ in proportion to $1/\sqrt{x}$, corresponding to the fact that
 the length scale in a weak magnetic field is given by the magnetic length $\sqrt{\hbar/(eB)}$.
 We have the same feature in $x=1$ symmetrically. At $x=1/2$, we notice that the $\xi(x)$ diverges only in the negative gaps of the subcells,
while it remains finite in the positive gaps. This links to the fact that the negative gaps close while the positive gaps are open
at $x=1/2$. For the main spectrum and any subcells centered at $E=0$,
the positive and negative gaps both closes at $x=1/2$ at the same time because of the electron-hole symmetry, 
and the localization lengths both diverge accordingly.
In Fig.\ \ref{fig:fractal}(b), the negative gaps of $R_{0}$ and $C_{0}R_{0}$ seem to follow a different curve 
from the rest in the limit of $x\to 0$, and this is related to the fact that the gap is approaching 
the Dirac point of the magnetic Bloch band at $E=0$.
The origin of the different scaling nature is argued in Supplemental Information.

\paragraph{Magnetic moment}


Generally, the in-gap defect states in a time-reversal-symmetry broken system
are accompanied by an orbital current circulation \cite{PhysRevB.95.115434,PhysRevB.101.245132}.
Here we find that the magnetic moment created by the orbital current of the defect states in our system is 
precisely related to  the gradient of fractal defect states on the Hofstadter diagram.
The local electric current $J_{nm}$ from site $m$ to site $n$ is calculated by
\begin{equation}
\label{eq:current}
J_{nm}=i \frac{(-e)t}{\hbar}(e^{i \theta _{mn}}\psi^*_n\psi_m-{\rm c.c.}).
\end{equation}
The right column in Fig.\ \ref{fig:amplitude} shows that the local current calculated for the defect states (i), (ii) and (iii).
 The thickness of and the size of arrows is proportional to the absolute value of current. 
Here the current rotates in the clockwise direction, i.e., it has a negative magnetic moment. Actually, the direction of the current synchronizes with the gradient of the defect energy level in the Hofstadter diagram,
 because the orbital magnetic moment is given by $m = -dE/dB$ (see Supplementary Information for the proof). 
In other words, a defect state in the Hofstadter butterfly precisely tunes its own current circulation, in such a way that the energy level stays inside the fractal gap in changing magnetic field.

\begin{figure}
    \centering
    \includegraphics[width=\linewidth]{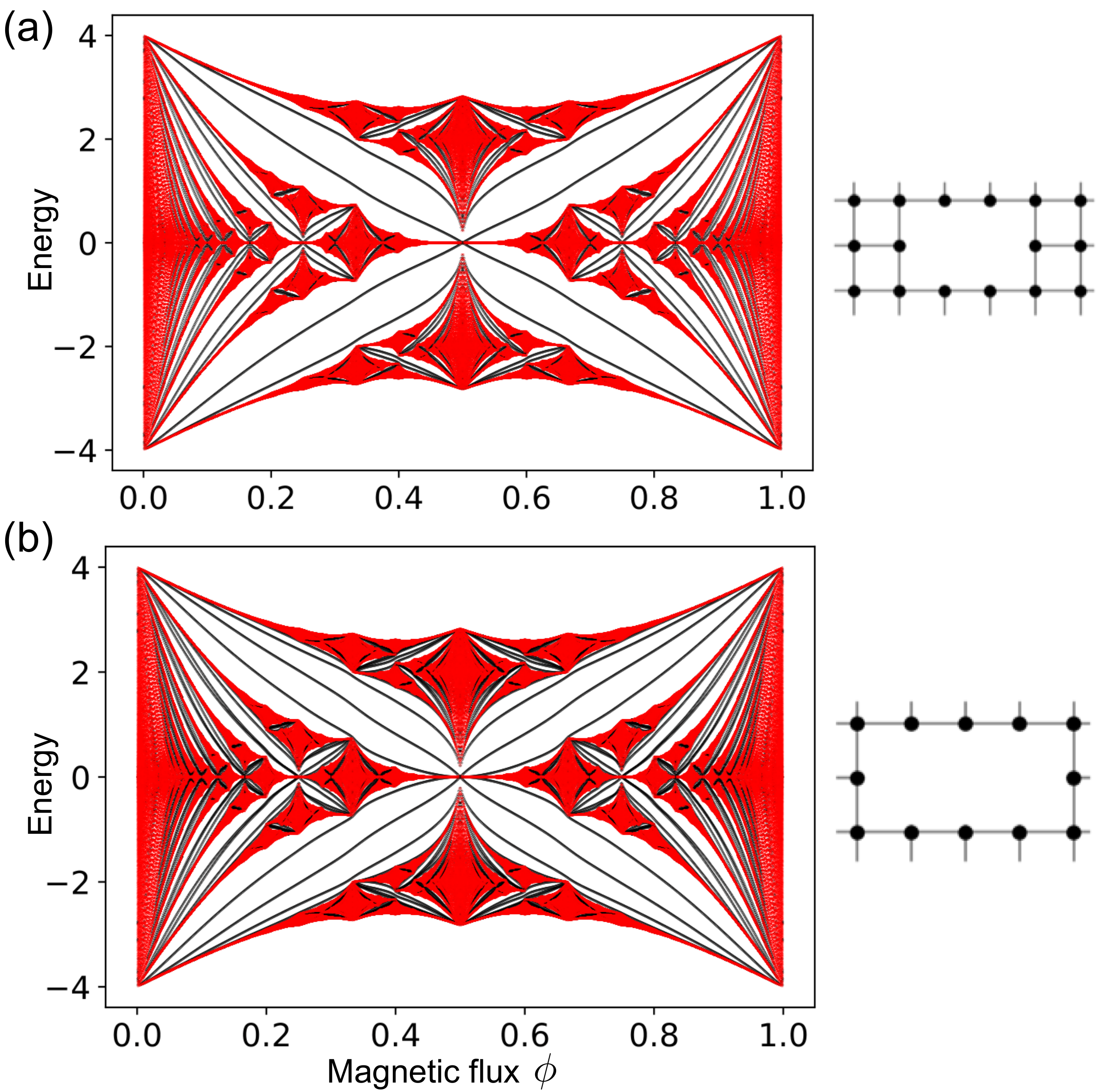}
    \caption{Energy spectra of $30 \times 30$ superlattice with
    (a) consecutive two-point defect and (b) consecutive three-point defect.}
    \label{fig:multi}
\end{figure}

\paragraph{Multi-point vacancies}

We also consider various multi-point vacancies in Fig.~\ref{fig:multi}; (a) consecutive two-point defect and (b) three-point defect. 
The corresponding spectra of $30 \times 30$ superlattice
are shown in Fig.~\ref{fig:multi}, 
where we observe that the number of defect states in every single fractal gap matches that of the missing sites in the defect.
When the number of missing atoms becomes larger, we expect that more and more defect levels fill in the energy gaps, and they eventually form quantum Hall edge states circulating around the hole. The fractal defect levels can be regarded as the quantized version of the quantum Hall edge state in the atomic limit. Therefore the emergence of the defect levels in the fractal gaps may be viewed as a sort of the bulk-edge correspondence \cite{PhysRevLett.71.3697,PhysRevB.48.11851} requiring the existence of the edge states in a bulk gap with a non-zero Chern number.
However, it should also be noted that the number of the defect states in each gap is not at all related to the Chern number, but it just coincides with the number of missing sites for any gaps.

\paragraph{Conclusion and Discussions} 

We have reported that the states localized around the defects fractally appear in every single band gap of the Hofstadter butterfly.
The defect states in different energy gaps have all different length scale in the spatial decay,
while they follow a universal curve after the appropriate fractal scaling.
Each defect state has its own characteristic magnetic moment, which is
exactly linked to the gradient of the corresponding bulk energy gap in the Hofstadter diagram.



While the previous observations of Hofstadter butterfly have been mainly conducted by spectroscopic/transport measurements of the energy gap structure, our work provides a powerful method to observe the fractal nature in the wavefunction by measuring the spatial decay of the defect states using scanning tunneling spectroscopy.

\paragraph{Acknowledgement} We thank Kin-ya Oda for fruitful discussions. This work was supported in part by Grant-in-Aid for JSPS Research Fellow, No. JP19J20559, JP19J11073, JSPS KAKENHI Grant Number JP20H01840 and JP20H00127
and by JST CREST Grant Number JPMJCR20T3, Japan.

\bibliographystyle{apsrev4-1.bst}
\bibliography{ref}



\clearpage
\onecolumngrid
\begin{center}
  \textbf{\large Supplemental Information for: Fractal defect states 
in the Hofstadter butterfly}\\[.2cm]
  Yoshiyuki Matsuki,$^{1}$ Kazuki Ikeda,$^{1}$ and Mikito Koshino$^1$\\[.1cm]
  {\itshape ${}^1$Department of Physics, Osaka University, Toyonaka, Osaka 560-0043, Japan\\}
\end{center}
\twocolumngrid

\section{The origin of unusual scaling curves of localization length}
In the main article, we demonstrated that the localization length of defect state in each minigap of the Hofstadter butterfly 
is approximately described by a single universal function when it is appropriately scaled.
However, there are some exceptional gaps in the spectrum, where the localization length follows a different curve.

\begin{figure}[H]
    \centering
    \includegraphics[width=\linewidth]{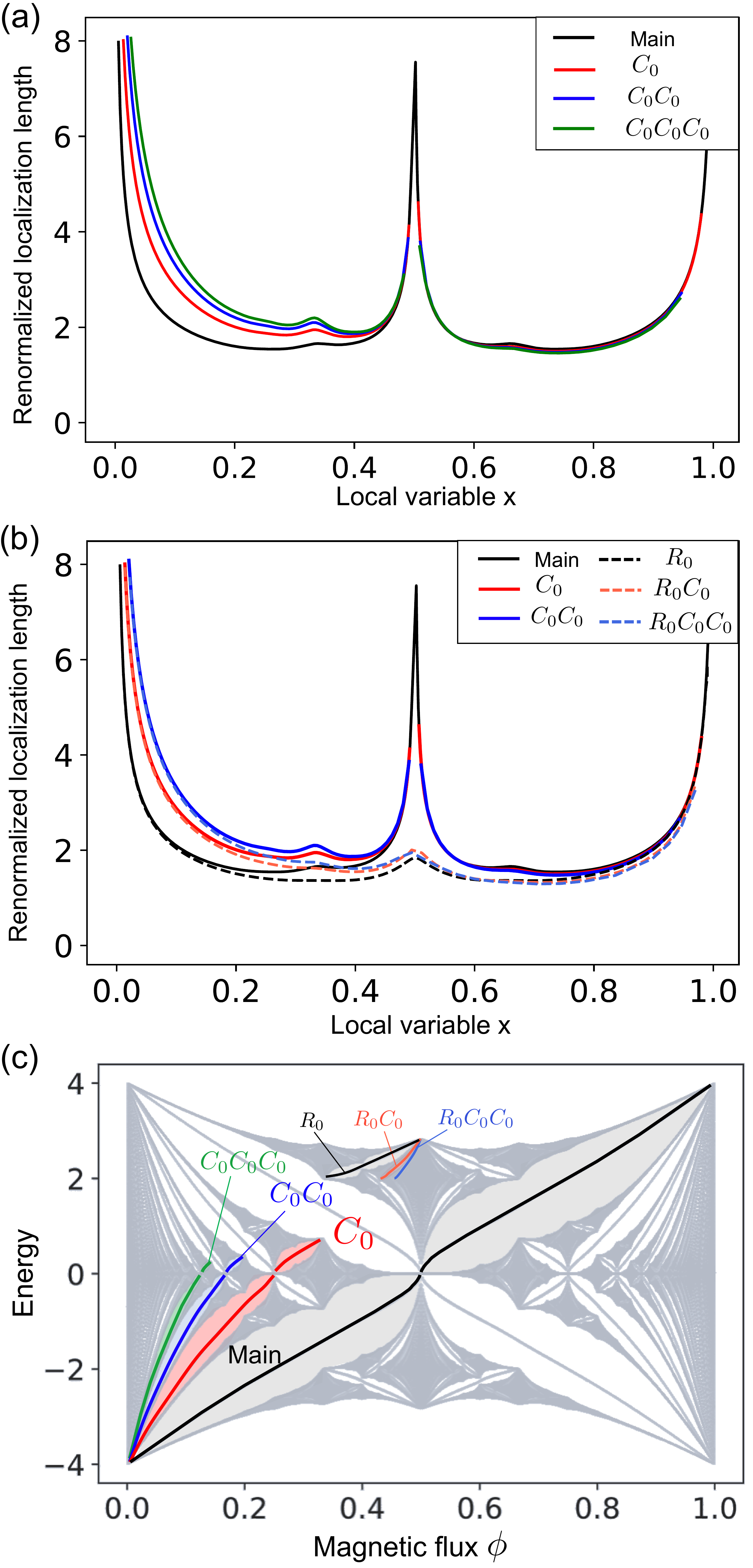}
    \caption{
 (a) Renormalized localization length of the positive principal gap of the main cell, $C_{0}$,
$C_{0}C_{0}$ and $C_{0}C_{0}C_{0}$ subcells
[$\xi$, $(\phi^{C_{0}}/x)\xi^{C_{0}}$, $(\phi^{C_{0}C_{0}}/x)\xi^{C_{0}C_{0}}$, $(\phi^{C_{0}C_{0}C_{0}}/x)\xi^{C_{0}C_{0}C_{0}}$, respectively] as functions of the local variable $x$. (b) 
The same plot as (a), with dashed lines indicating the renormalized localization length for 
$R_{0}$, $R_{0}C_{0}$ and $R_{0}C_{0}C_{0}$ subcells
[$\phi^{R_{0}}\xi^{R_{0}}$, $(\phi^{R_{0}C_{0}}\phi^{C_{0}}/x)\xi^{R_{0}C_{0}}$, $(\phi^{R_{0}C_{0}C_{0}}\phi^{C_{0}C_{0}}/x)\xi^{R_{0}C_{0}C_{0}}$, respectively]. (c) The positive principal gaps of main cell, $C_{0}$, $C_0C_0$ and $C_0C_0C_0$ subcells, which correspond to the first, second, third and forth Landau-level gaps in $\phi \to 0$, respectively. Similarly, the positive principal gap of $R_0C_0$ and $R_0C_0C_0$ subcells correspond to the second and third Landau-level gaps, respectively,
of the magnetic Bloch band in $\phi \to 1/2$.
}
\label{fig:C0}
\end{figure}

An exceptional scaling behavior occurs in the principal gaps of any subcells ending with $C_{0}$,
i.e., $XY\cdots C_{0}$.
Figure \ref{fig:C0}(a) shows the renormalized localization length [Eqs.\ (6) and (7) of the main text] 
of the positive principal gap of the main cell, $C_{0}$, $C_{0}C_{0}$ and $C_{0}C_{0}C_{0}$ subcells
[$\xi$, $(\phi^{C_{0}}/x)\xi^{C_{0}}$, $(\phi^{C_{0}C_{0}}/x)\xi^{C_{0}C_{0}}$, $(\phi^{C_{0}C_{0}C_{0}}/x)\xi^{C_{0}C_{0}C_{0}}$] as functions of the local variable $x$.
We observe that the curves slightly shift in relative to each other in the limit of $x\to 0$, whereas they precisely match in $x \gtrsim 1/2$. 

The reason for the deviation can be understood as follows. 
In the weak magnetic field limit,
the principal gaps of $C_{0}$, $C_{0}C_{0}$ and $C_{0}C_{0}C_{0}$ subcell
correspond to the second, third and fourth lowest gaps
of the Landau-level spectrum, respectively [Fig.\ \ref{fig:C0}(c)].
This is in contrast to any subcells NOT ending with $C_{0}$, where the principal gap definitely connects
to the lowest gap (the gap just above the lowest Landau level) in the magnetic Bloch subband at $x=0$.
The characteristic length scale of the $n$-th Landau level wavefunction $\varphi_n$ is given by $\sqrt{n+1/2} \,l_B$, 
where $l_B = \sqrt{\hbar/(eB)}$ is the magnetic length.
Accordingly, the defect-localized state which exists between the Landau levels $n-1$ and $n$
should have the length scale of the order of $\sim \sqrt{n}\,l_B$,
and the dependence on $n$ results in the different scaling curves.
Indeed, the localization lengths of the main cell, $C_{0}$, $C_{0}C_{0}$,and $C_{0}C_{0}C_{0}$ 
in the region of $x<0.01$ in Fig.\ \ref{fig:C0}(a)
are very well (within 1\%) fitted by 

\begin{equation}
\xi \approx 1.72\sqrt{n-0.223} \, l_B,
\label{eq_xi_n}
\end{equation}
with $n =1,2,3$ and $4$, respectively.
Here we note that $l_B \approx 1/\sqrt{2\pi x}$ (in units of the lattice constant) in this region.


Similarly, the same scaling rule is applicable to $XY\cdots Z(C_{0})^n$ subcells.
In Fig.\ \ref{fig:C0}(b), we plot the renormalized localization length of
 the principal gaps of $R_{0}$, $R_{0}C_{0}$ and $R_{0}C_{0}C_{0}$ subcells by dashed curves.
In the limit $x\to 0$, the three curves perfectly match with those of the main cell, $C_{0}$ and $C_{0}C_{0}$, respectively (solid curves).


Another exceptional case occurs in a gap connected to the Dirac point of the magnetic Bloch band, such as the negative principal gaps of $R_0$ and $C_0R_0$ mentioned in the main text. If we take $R_0$ subcell, for instance, the gap leads to the point of $\phi=1/2$ and $E=0$ of the main diagram in the limit of $x\to 0$, where a pair of magnetic Bloch bands are touching just like in graphene due to the electron-hole symmetry of the model. Now the wavefunction of the Landau level $n$ in graphene is composed of the $\varphi_{n-1}$ and $\varphi_{n}$ at different sublattices [{\ref{JPSJ67_2421-2429_1998}},{\ref{PhysRevB.65.245420}}], so that its length scale is just in the middle of those of $(n-1)$-th and $n$-th Landau levels in the conventional massive electron.
Accordingly, the defect-localized states should have the intermediate localization length compared to the massive system.
Indeed, the localization length of defect states in $R_{0}$ cell for small $x$ is given by $\xi \approx 1.98\, l_B$, which is between the values of $n=1$ and $2$ in Eq. (\ref{eq_xi_n}).
In Fig.\ 4(b) in the main text, we see that the $C_0R_0$ subcell also follows the same curves as $R_0$.
This is because its negative principal gap leads to the point of $\phi=1/4$ and $E=0$, which is also the Dirac point.

\section{The gradient of defect states and the magnetic moment} 
We prove that the gradient of a defect energy level in the Hofstadter diagram coincides with
the magnetic moment created by the local electric current. 
For this purpose we show that the magnetic moment $\vector{m}$ obeys Eq.~\eqref{eq:magnet} in the presence of a generic potential $V(\vector{r})$ in quantum mechanics. To show this formula we consider the system described by the Hamiltonian,
\begin{equation}
H_{0}=\frac{1}{2m}(-i\vector{\nabla}+e\vector{A})^2+V(\vector{r})~~(\hbar=1).
\end{equation}
Considering a small change in the magnetic field, we obtain the perturbed Hamiltonian
\begin{eqnarray}
H&&=\frac{1}{2m}(-i\vector{\nabla}+e(\vector{A}+\delta \vector{A}))^2+V(\vector{r})\\
&&=H_{0}-\frac{1}{2}(\vector{J}\cdot \delta\vector{A}+\delta \vector{A}\cdot \vector{J})\equiv H_{0}+\delta H,
\end{eqnarray}
where $\vector{J}$ is the current operator
\begin{equation}
\vector{J}=(-e)i[\vector{r},H_{0}]=(-e)\frac{-i\vector{\nabla}+e\vector{A}}{m}.
\end{equation}
The variation of the energy $\delta E$ within the first order perturbation is
\begin{equation}
\label{eq:var}
\delta E=\bra{\psi_{0}}\delta H\ket{\psi_{0}}=-\frac{1}{2}\bra{\psi_{0}}(\vector{J}\cdot \delta\vector{A}+\delta \vector{A}\cdot \vector{J})\ket{\psi_{0}},
\end{equation}
where $\psi_{0}$ is the eigenfunction of $H_{0}$. The first term in the most right hand side in (\ref{eq:var}) is evaluated as follows:
\begin{align}
\begin{aligned}
&\bra{\psi_{0}}(\vector{J}\cdot \delta\vector{A})\ket{\psi_{0}} \\ 
&=\int d\vector{r} d\vector{r}^{\prime}\psi^{\ast}_{0}(\vector{r})\vector{J}(\vector{r})\delta(\vector{r}-\vector{r}^{\prime})\psi_{0}(\vector{r})\cdot \delta\vector{A}(\vector{r}^{\prime}).
\end{aligned}    
\end{align}

We can evaluate the second term in a similar way, and finally obtain
\begin{eqnarray}
\delta E=\int d\vector{r}^{\prime} \vector{j}(\vector{r}^{\prime})\cdot \delta\vector{A}(\vector{r}^{\prime}).
\end{eqnarray}
Here we used the local current operator $\vector{j}$,
\begin{equation}
\vector{j}(\vector{r}^{\prime})\equiv-\frac{1}{2}\int d\vector{r}\psi^{\ast}_{0}(\vector{r})(\vector{J}(\vector{r})\delta (\vector{r}-\vector{r}^{\prime})+\delta (\vector{r}-\vector{r}^{\prime})\vector{J}(\vector{r}))\psi_{0}(\vector{r}).
\end{equation}
By using the expression of the magnetizing current $\vector{j}=\vector{\nabla} \times \vector{m}$, one can find that the magnetic moment $\vector{m}$ obeys
\begin{equation}
\label{eq:magnet}
\vector{m}=-\frac{dE}{d\vector{B}}. 
\end{equation}

\section{The relationship between the distance of defect sites and the fractal energy spectrum}
\begin{figure}[H]
    \centering
    \includegraphics[width=\linewidth]{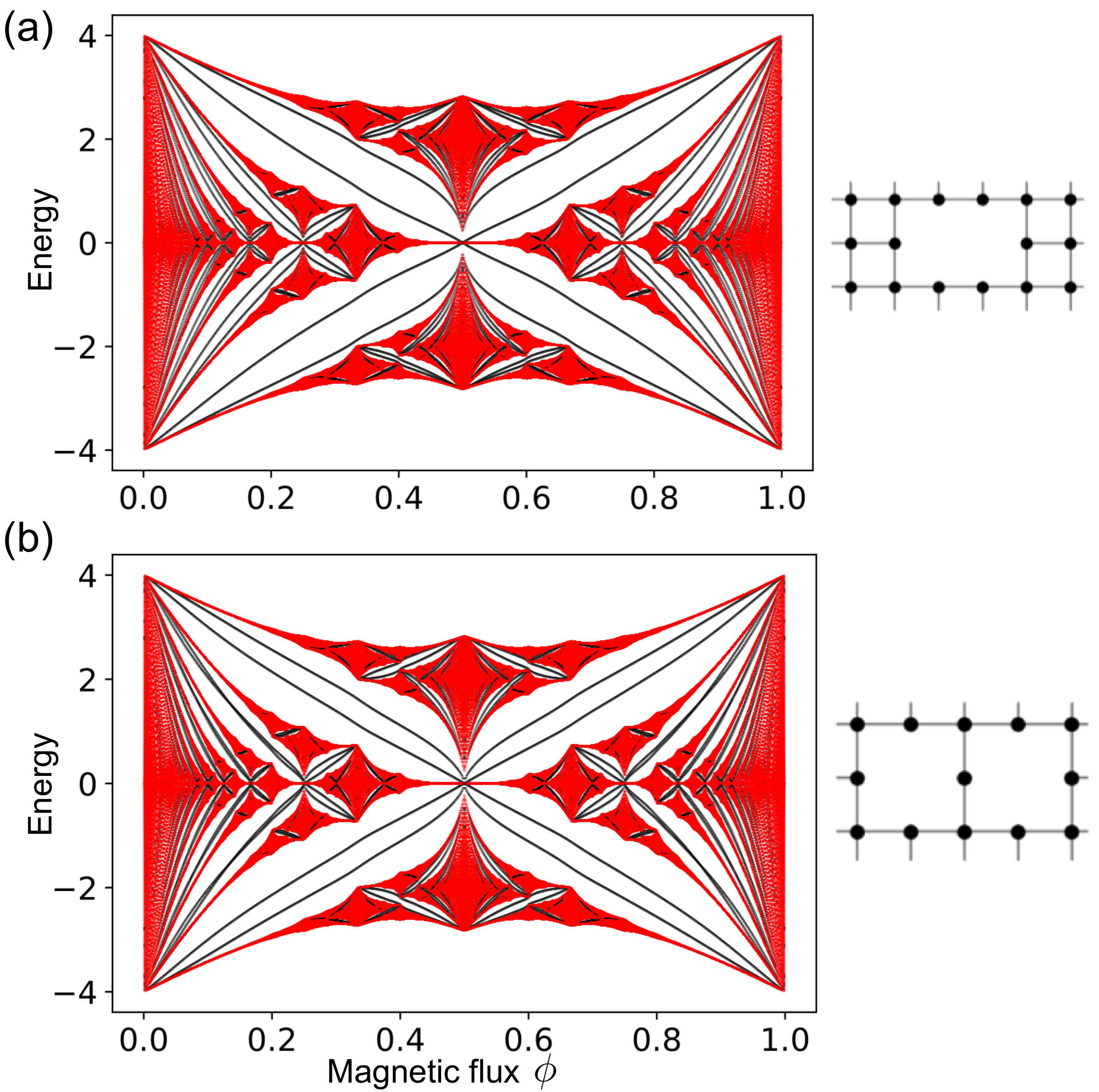}
    \caption{(Color online) Energy spectra of $30 \times 30$ superlattice with
    (a) consecutive two-point defect and (b) two split defects.}
    \label{fig:twodefect}
    \end{figure}

Here we discuss the relationship between the distances of vacancy sites and the fractal energy levels in the Hofstadter diagram. For this purpose, we consider the two-point vacancies illustrated in Fig.\ref{fig:twodefect}: (a) consecutive two-point defect, (b) two split defects. The corresponding spectra of $30 \times 30$ superlattice are shown in Fig.\ref{fig:twodefect}(a) and (b). By comparing Figs.\ref{fig:twodefect}(a) and (b), we notice that the two defect energy levels get closer as the distance between the two defect sites gets further away. This is a consequence of the hybridization of the defect states of two single vacancy sites, where the coupling strength exponentially decreases as the distance increases. A similar effect is also found in defect states in graphene [\ref{PhysRevB.78.125402}]. We see the same tendency consistently in all the fractal gaps.

\subsection{Fractal Structure and Wavefunction on a honeycomb lattice with a defect}
\begin{figure}[H]
    \centering
    \includegraphics[width=\linewidth]{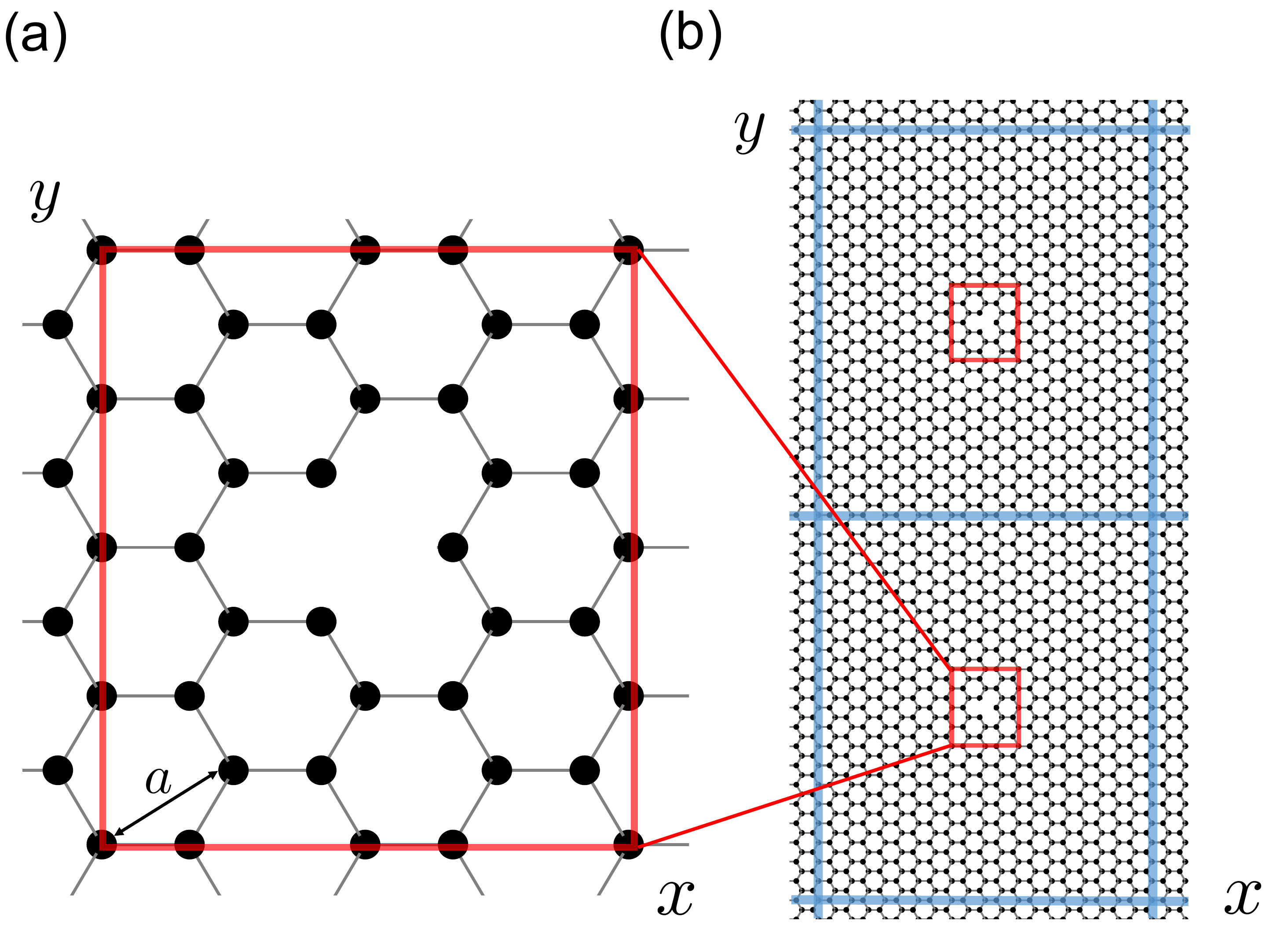}
    \caption{(Color online) Honeycomb lattice with a single point vacancy and (b) its periodic superlattice with $20\times 20$ supercel}
    \label{fig:honeycomblat}
    \end{figure}
Here we consider defect-localized states on a honeycomb superlattice with a single-site defect (Fig. \ref{fig:honeycomblat}). We assume that the system is periodic with $N\times N$ supercell and every single supercell includes a single vacancy site. Now we take the super period $N=20$, which is taken to be sufficiently large to avoid interference between defect-localized states in neighboring cells. The Bloch electron wavefunctions obey the magnetic Bloch condition:
\begin{eqnarray}
\psi(\vector{r}+\vector{L}_{1})&=&e^{ik_{x}L_{1}} e^{-i2\pi e BL_{1}y}\psi(\vector{r}),\\
\psi(\vector{r}+\vector{L}_{2})&=&e^{ik_{y}L_{2}}\psi(\vector{r}),
\end{eqnarray}
where $\vector{L}_{1}=(\sqrt{3}Na/2,0)$ and $\vector{L}_{2}=(0,Na)$ are the primitive lattice vectors of the magnetic unit cell.

\begin{figure}[H]
    \centering
    \includegraphics[width=0.92\linewidth]{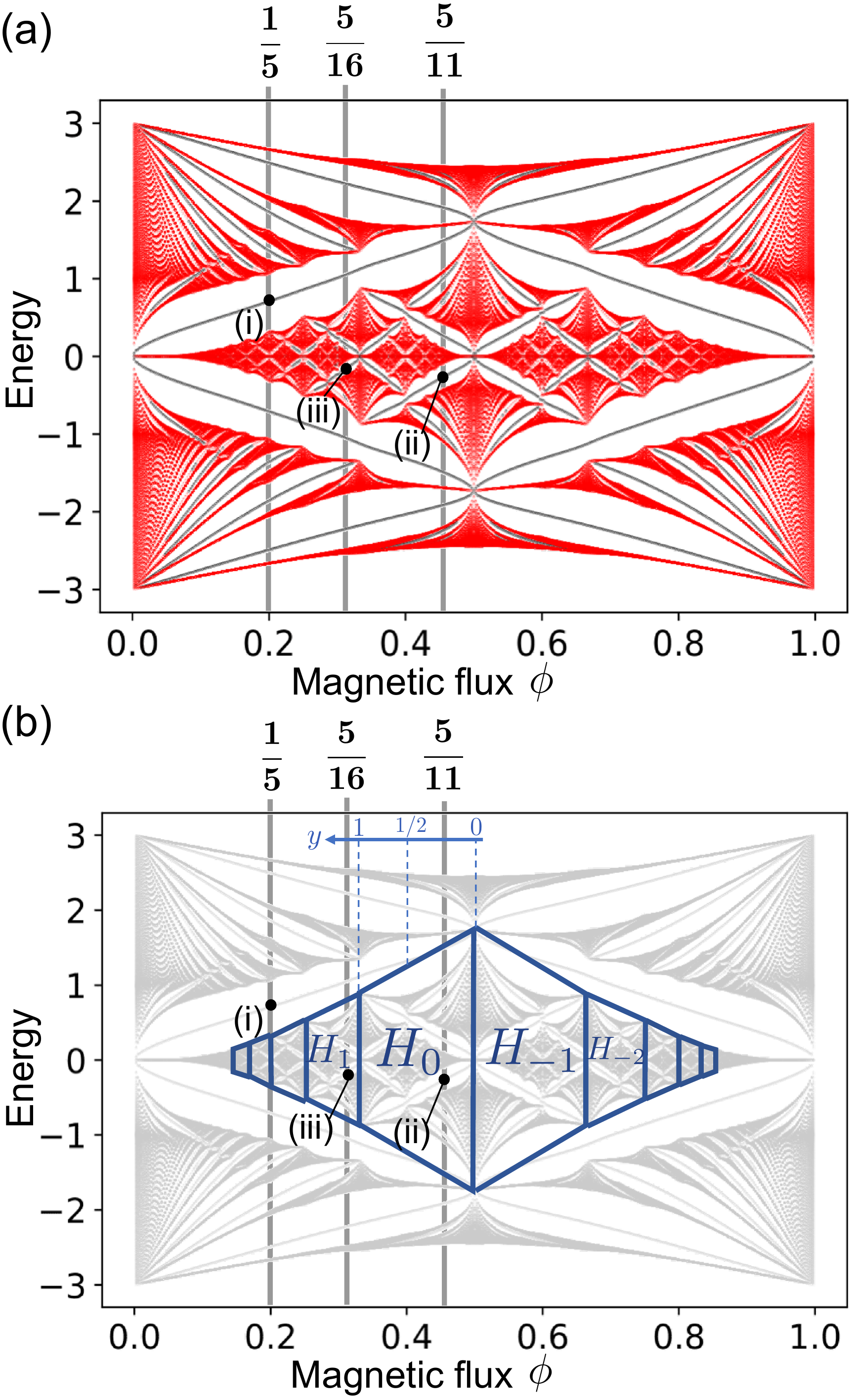}
    \caption{(a) Energy spectrum of $20 \times 20$ superlattice with a single-site defect, which is plotted against the magnetic flux $\phi$. The red and black dots represent the bulks states and the defect-localized states, respectively. The labels (i), (ii) and (iii) correspond to the wavefunctions shown in Fig.\ \ref{fig:honeycombamplitude}. (b) Subcell decomposition of the Hofstadter butterfly. The states (i), (ii) and (iii) belong to the positive gradient principal gaps of the main spectrum, $H_0$ and $H_1$, respectively.}
    \label{fig:honeycombbutterfly}
    \end{figure}

FIG.\ref{fig:honeycombbutterfly}(a) shows the energy spectrum of $20\times 20$ superlattice in a honeycomb lattice with a single defect, plotted against the magnetic flux $\phi$. The red and black dots represent the bulk states and the defect-localized states, respectively. Here the defect-localized states are identified by the condition that the wave amplitude within $2\sqrt{3}a$ from the defect point is more than $50\%$ of the total amplitude. We observe that a defect level exists in every single gap, indicating that the spectrum of the defect states inherits the nested fractal structure of the Hofstadter butterfly, same as the square lattice case.

\begin{figure}[H]
    \centering
    \includegraphics[width=\linewidth]{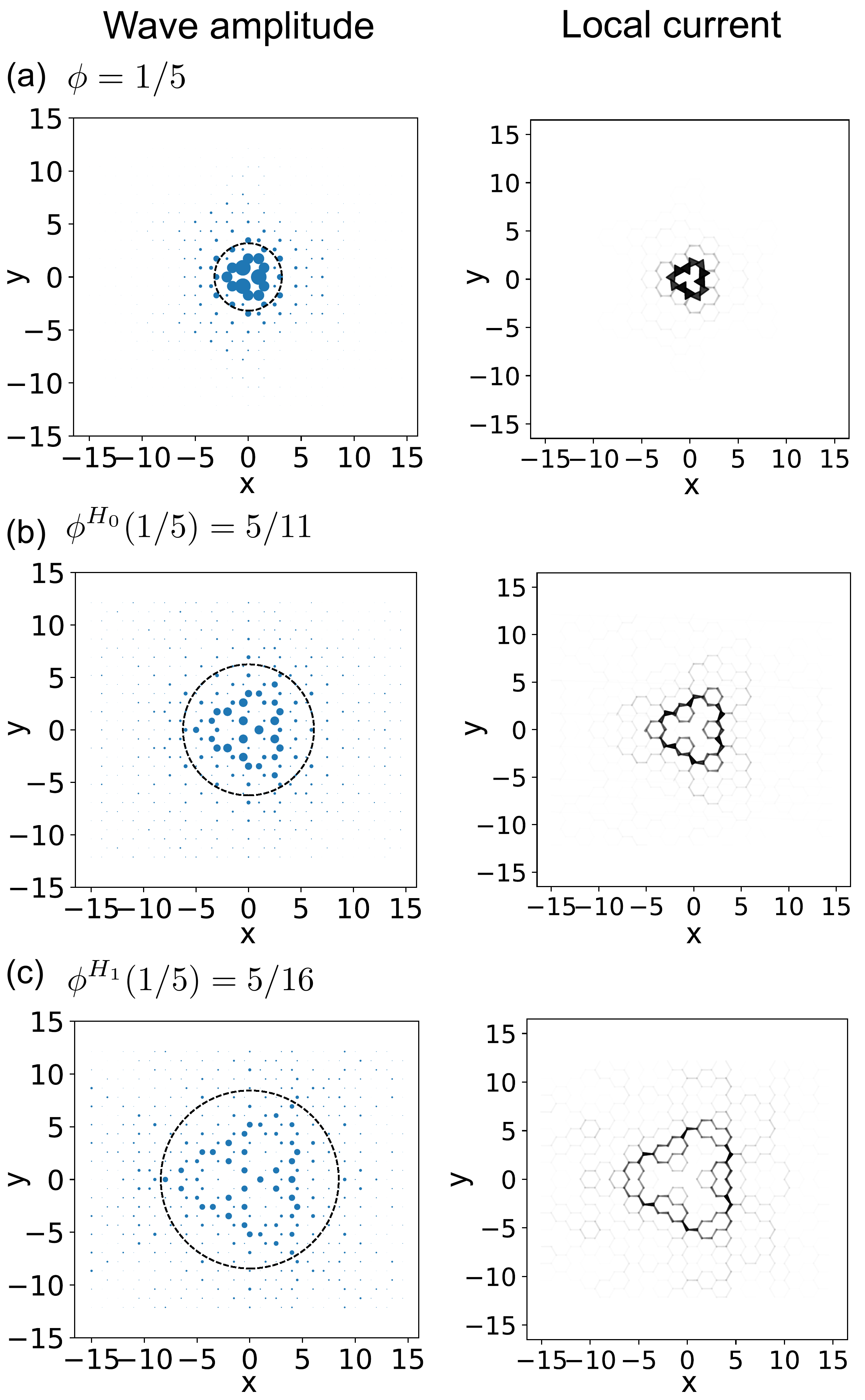}
    \caption{Distribution of the wave amplitude (left) and the local electric current (right) of the defect states (i), (ii) and (iii), which are indicated in Fig.\ref{fig:honeycombbutterfly}. The dots size in the left figures represents the magnitude of wavefunction, and thickness and color depth of the arrows in the right figures represent the local current amplitude.}
    \label{fig:honeycombamplitude}
    \end{figure}

To compare uniformly the localization lengths in different fractal generation gaps, we again define the subcell decomposition and the local variable $y$ in a honeycomb lattice. As shown in Fig.\ref{fig:honeycombbutterfly}(b), the main energy spectrum in $0\leq \phi \leq 1$ includes the self-similar structure subcell $H_{n}$ $(n=0,\pm1,\cdots)$. In fact, the gap structure of each subcell plotted against the local variable $y$ $(0\leq y \leq 1)$ is identical to that of the main spectrum plotted against the magnetic flux $\phi$ $(0\leq \phi \leq 1)$. The relationship between the global variable $\phi$ and local variable $y$ is 
\begin{equation}
\label{eq:localvar_honeycomb}
\phi^{H_{n}}(y)=(n+y+2)^{-1}.
\end{equation}
In the main spectrum, the global variable $\phi$ corresponds to the local variable $y$ since the main spectrum can be viewed as a single subcell. In the following, we show the amplitude and the electric current of the corresponding defects states of different subcells at the same local variable $y$.

The left panels in Fig.\ref{fig:honeycombamplitude} show the distribution of the wave amplitude of three defect levels (i), (ii) and (iii) indicated in Fig. \ref{fig:honeycombbutterfly}, which are taken from the positive principal gaps of the main spectrum, $H_{0}$ and $H_{1}$, respectively, with the same local variable $y=1/5$. From the equation (\ref{eq:localvar_honeycomb}), the global variables $\phi$ for (i), (ii) and (iii) are $1/5$, $\phi^{H_{0}}(1/5) = 5/11$ and $\phi^{H_{1}}(1/5) = 5/16$, respectively. We can observe that the defect wavefunctions localize around the defect with different length scales. 

Moreover, the right panels in Fig.\ref{fig:honeycombamplitude} represent that the local current calculated for the defect states (i), (ii) and (iii). As discussed above, the magnetic moment created by the local current coincides with the gradient on the Hofstadter diagram (Fig.\ref{fig:honeycombbutterfly}). Actually, the current rotates in the clockwise direction, and the direction of the current synchronizes with the gradient of the defect energy level in the Hofstadter diagram. The same properties as the square lattice case (the localization around the defect, the fractality of localization length, and the correspondence between magnetic moment and gradient of defect states) are observed in a honeycomb lattice case. 

\begin{center}
    \rule{0.35cm}{0.5pt}\rule[-0.2pt]{0.35cm}{0.9pt}\rule[-0.25pt]{0.3cm}{1.0pt}\rule[-0.45pt]{0.3cm}{1.4pt}\rule[-0.45pt]{0.3cm}{1.4pt}\rule[-0.25pt]{0.3cm}{1.0pt}\rule[-0.2pt]{0.35cm}{0.9pt}\rule{0.35cm}{0.5pt}
\end{center}
\renewcommand{\theenumi}{A\arabic{enumi}}
\renewcommand{\labelenumi}{[\theenumi]}
\begin{enumerate}
    \item\small{N. H. Shon and T. Ando, Journal of the Physical Society of Japan, {\bf 67}, 2421-2429 (1998).}\label{JPSJ67_2421-2429_1998} 
    
    \item\small{Y. Zheng and T. Ando, Phys. Rev. B {\bf 65}, 245420 (2002).} \label{PhysRevB.65.245420}
    
    \item\small{A. L. C. Pereira and P. A. Schulz, Phys. Rev. B {\bf 78}, 125402 (2008).}\label{PhysRevB.78.125402}
\end{enumerate}


 \end{document}